\documentclass{osa-article}

\journal{osajournal}

\articletype{Research Article}

\usepackage{physics}

\begin{document}

\title{Selective active resonance tuning for multi-mode nonlinear photonic cavities}

\author{Alan D. Logan\authormark{1,$\dagger$,*}, Nicholas S. Yama\authormark{2,$\dagger$}, and Kai-Mei C. Fu\authormark{1,2,3}}

\address{
\authormark{1} Department of Physics, University of Washington, Seattle WA 98195, USA\\
\authormark{2}Department of Electrical and Computing Engineering, University of Washington, Seattle WA 98195, USA\\
\authormark{3} Physical Sciences Division, Pacific Northwest National Laboratory, Richland, Washington 99352, USA\\
\authormark{$\dagger$} These authors contributed equally to this work.
}
\email{\authormark{*}adlogan@uw.edu}

\begin{abstract*} 
Resonant enhancement of nonlinear photonic processes is critical for the scalability of applications such as long-distance entanglement generation.
To implement nonlinear resonant enhancement, multiple resonator modes must be individually tuned onto a precise set of process wavelengths, which requires multiple linearly-independent tuning methods.
Using coupled auxiliary resonators to indirectly tune modes in a multi-resonant nonlinear cavity is particularly attractive because it allows the extension of a single physical tuning mechanism, such as thermal tuning, to provide the required independent controls.
Here we model and simulate the performance and tradeoffs of a coupled-resonator tuning scheme which uses auxiliary resonators to tune specific modes of a multi-resonant nonlinear process.
Our analysis determines the tuning bandwidth for steady-state mode field intensity can significantly exceed the inter-cavity coupling rate $g$ if the total quality factor of the auxiliary resonator is higher than the multi-mode main resonator.
Consequently, over-coupling a nonlinear resonator mode to improve the maximum efficiency of a frequency conversion process will simultaneously expand the auxiliary resonator tuning bandwidth for that mode, indicating a natural compatibility with this tuning scheme.
We apply the model to an existing small-diameter triply-resonant ring resonator design and find that a tuning bandwidth of $136\, \mathrm{GHz} \approx 1.1$\,nm can be attained for a mode in the telecom band while limiting excess scattering losses to a quality factor of $10^6$.
Such range would span the distribution of inhomogeneously broadened quantum emitter ensembles as well as resonator fabrication variations, indicating the potential for the auxiliary resonators to enable not only low-loss telecom conversion but also the generation of indistinguishable photons in a quantum network.
\end{abstract*}

\section{Introduction}

Integrated photonic resonators can greatly enhance the efficiency of nonlinear processes, with applications in frequency conversion~\cite{lu2019ppln_ring_shg, guo2016strong, wang2021degenerate, ye2020triple, chen2019ultra, samblowski2014conversion, Lin2014shg_continuous, logan2023sfg, wilson2020gap_comb} and photon-pair production~\cite{azzini2012resonator_sfwm,wakabayashi2015si_resonator_photon_pair,fortsch2013disk_ln_spdc}.
In such devices, optimal performance occurs when the resonator has a set of high-quality-factor modes (collectively satisfying phase matching conditions) which are each centered at one of the participant frequencies of the targeted process.
However, achieving this multiple-resonance condition for any set of frequencies --- let alone for a specific set --- requires precision beyond what is attainable by current fabrication processes.
Even if such precision were attainable, transient fluctuations in temperature or humidity, or gradual material relaxation, will inevitably disrupt the multi-resonance structure.
Consequently, practical implementations require a means of dynamically tuning the resonance frequencies in order to compensate for both fabricated and dynamic variations.

Photonic resonators can be tuned by a variety of mechanisms, including static methods such as post-fabrication modification of the resonator~\cite{prorok2012trimming, schrauwen2008trimming, milosevic2018ion,moille2021postfab_etch} or cladding~\cite{thiel2022trimming, zhou2009athermalizing, atabaki2013accurate, spector2016localized, biryukova2020trimming, riemensberger2012ald_trimming, atabaki2013trim_cladding_etch}, or active methods via temperature, electro-optic effect, or mechanical strain~\cite{lu2019ppln_ring_shg, guo2016strong, logan2023sfg, bruch201817, logan2018shg, chen2019ultra, samblowski2014conversion, Lin2017nonlinear_WGM_rev, Lin2014shg_continuous, fortsch2013disk_ln_spdc}.
Such methods have been shown to enable double or even triple resonance provided the tuning mechanism has a different effect on each mode~\cite{Lin2017nonlinear_WGM_rev,logan2023sfg,bruch201817,lu2019ppln_ring_shg}.
However, these techniques employ blanket alterations of the multi-mode structure which necessarily affect all resonances simultaneously~\cite{Lu2020splitting}.
Consequently, the particular wavelengths for which the multi-resonance condition can be achieved are still determined by the static device dimensions, including fabrication imperfections.
Alternative, mode-targeted wavelength trimming methods~\cite{Lu2020splitting, Lu2014selective_splitting, deGoede2021patterned_splitting} have also been demonstrated and could be used to achieve multi-resonance at specific frequencies, though these corrections are static and cannot be used to compensate for dynamic variations.
An ideal tuning mechanism should combine the features of these paradigms, enabling dynamic tuning of each mode individually.
Such control would not only increase tolerance to environmental and fabrication variances, but would also find use in quantum networks to correct the spectral inhomogeneity of optically active defects~\cite{levonian2022distinguishable, weber2019twophoton}.

\begin{figure}[t]
    \centering
    \includegraphics[width=\textwidth]{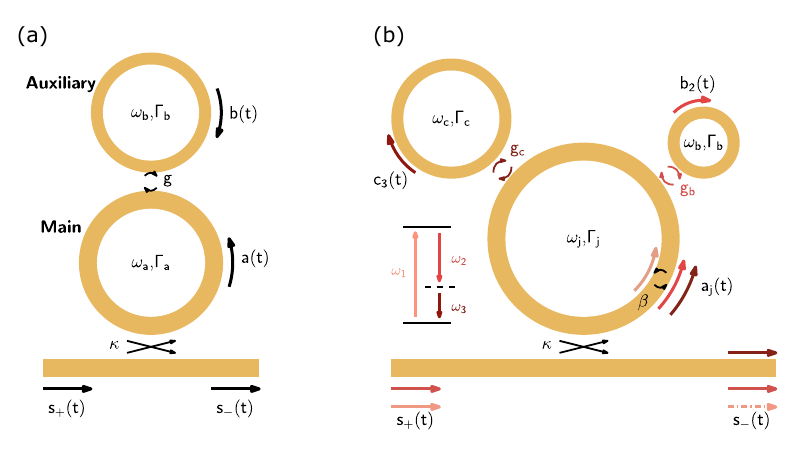}
    \caption{
    (a) Diagram of a main ring resonator with frequency $\omega_a$ and loss $\Gamma_a$ is coupled to an auxiliary cavity of frequency $\omega_b$ and loss $\Gamma_b$ at a rate of $|g|$.
    This system is modeled in Section 2.
    (b) A potential layout for sum/difference frequency generation using two auxiliary resonators.
    For simplicity, input/output coupling for all three main resonator modes is shown with a single waveguide.}
    Each auxiliary resonator can only couple to one of the frequencies $\omega_j$ in the main resonator.
    This system is modeled in Section 3.
    
    \label{fig:model}
\end{figure}

In this paper, we develop a complete description of an active tuning scheme for multi-resonant nonlinear resonators which uses coupled auxiliary resonators to independently tune the participant modes (Fig.~\ref{fig:model}).
Through selective coupling in the frequency domain, a targeted mode can be isolated within the auxiliary resonator and tuned independently.
While similar schemes have been explored experimentally~\cite{gentry2014fwm_auxres, souza2015coupled_ring_heaters, chew2010phc_mems_coupling, liu2021mzi_splitting_tuning}, we derive, in full generality, an analytical model for the auxiliary-resonator-based tuning mechanism based on a temporal coupled-mode theory description. 
It is shown that an individual mode in an overcoupled nonlinear resonator can be tuned on a scale exceeding the inter-cavity coupling rate $|g|$ --- which may already be significantly larger than the resonance linewidth --- without significant degradation of the conversion efficiency.
The formalism is directly applied to triply-resonant difference frequency conversion, in which we show there is negligible degradation of the conversion efficiency over several linewidths of detuning.
Practical considerations, such as the range of physically attainable inter-cavity coupling rates and quality factors are studied in the case of small-diameter gallium phosphide ring resonators.
For a main resonator with telecom-band intrinsic and coupling quality factors of $Q_i = 10^6$ and $Q_c = 10^5$ respectively, simulations find inter-cavity coupling rates as high as $|g|/2\pi=136$\,GHz are attainable without inducing significant excess losses, yielding a tuning bandwidth of approximately $64$ linewidths.
These calculations indicate that auxiliary resonator tuning may be the most promising tuning technology to realize high-yield multi-resonant devices at targeted operating frequencies.

\section{Orthogonal tuning by selective mode coupling}

\begin{figure}[t]
    \centering
    \includegraphics[width=\textwidth]{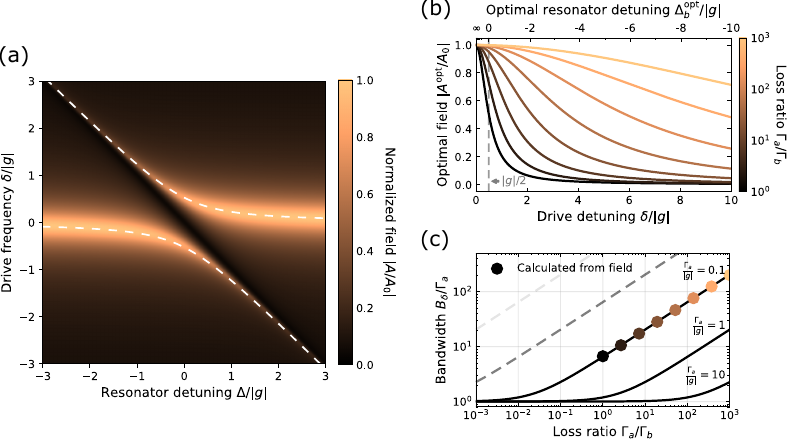}
    \caption{(a) Main cavity field amplitude as a function of drive and resonator detunings normalized to the resonantly driven cavity in the absence of an auxiliary ring.
    Dashed lines track the resonances given by $\Delta_b^{\text{opt}}(\delta)$ (Eq.~\eqref{eq:optimal_detuning}).
    Here $\Gamma_a/|g| = 0.5$ and $\Gamma_b/|g| = 0.1$.
    (b) Normalized field amplitudes along the resonance peaks as a function of drive detunings for different values of $\Gamma_a/\Gamma_b$ with $\Gamma_a/|g|=0.1$.
    (c) The 3-dB tuning bandwidth $B_\delta$ (Eq.~\eqref{eq:tuning_bandwidth}) as a function of $\Gamma_1/\Gamma_2$ for different $|\Gamma_a|$.
    Points are bandwidths obtained from fits the corresponding curves in (b).
    }
    \label{fig:two_rings_model}
\end{figure}

Auxiliary resonator tuning uses modifications of an auxiliary resonator structure to indirectly influence specific coupled modes in a main resonator while leaving other main resonator modes unperturbed.
In principle, the tuning scheme could be implemented using any resonator tuning mechanism to control the auxiliary resonator, including thermal or electro-optic tuning.
Many performance metrics, such as tuning speed and reversibility, are inherited from the chosen physical tuning mechanism.
In this section, a model of the auxiliary resonator tuning system is developed in order to describe behaviors that are inherent to the tuning scheme.

A single-mode auxiliary resonator system is shown in Fig.~\ref{fig:model}a. The main resonator supports a mode with complex field amplitude $a$ at an unperturbed frequency $\omega_a$, which is coupled to mode $b$ at frequency $\omega_b$ in the auxiliary resonator with an inter-cavity coupling rate $g$ as well as an input/output waveguide. The field evolution in this is system is described by the coupled-mode theory (CMT) model
\begin{subequations}
    \begin{align}
        \label{eq:aux_ring_model_a}
        \dv{a}{t} &= -i\left( \omega_a - i\frac{\Gamma_a}{2} \right) a - i \frac{g^*}{2} b + \sqrt{\kappa} s_{+}, \\
        \label{eq:aux_ring_model_b}
        \dv{b}{t} &= -i\left( \omega_{b} - i\frac{\Gamma_{b}}{2} \right) b - i \frac{g}{2} a, \\
        \label{eq:aux_ring_model_s}
        s_- &= -s_+ + \sqrt{\kappa} a,
    \end{align}
\end{subequations}
where $s_\pm$ are the incoming/outgoing waveguide field amplitudes (normalized so that $|s_\pm|^2$ is the input/output power) and $\kappa$ is the main-resonator-waveguide coupling rate.
The intrinsic ($\gamma_a$, $\gamma_b$) and coupling ($\kappa$) loss rates for each mode combine to form the total energy loss rates $\Gamma_a = \gamma_a + \kappa$ and $\Gamma_b = \gamma_b$.

The steady-state response to monotone driving at frequency $\omega$ is obtained by a Fourier transform: $a(t) \to A(\omega)$, $b(t) \to B(\omega)$, and $s_\pm(t)\to S_\pm(\omega)$.
The auxiliary resonator field $B$ can be eliminated from the resulting algebraic system of equations to yield a single resonator equation
\begin{equation}
    0 = - \left(i\delta_{\text{eff}} + \frac{\Gamma_{\text{eff}}}{2}\right) A + \sqrt{\kappa_j} S_{+}
\end{equation}
where 
\begin{align}
    \label{eq:effective_eval_1}
    \delta_{\text{eff}} &= \delta - \frac{|g|^2}{4} \frac{(\Delta_{b} + \delta)}{(\Delta_{b} + \delta)^2 + \Gamma_{b}^2/4}, \\
    \label{eq:effective_eval_2}
    \Gamma_{\text{eff}} &= \Gamma_a + \frac{|g|^2}{4} \frac{\Gamma_{b}}{(\Delta_{b} + \delta)^2 + \Gamma_{b}^2/4},
\end{align}
are the effective detuning and loss of the combined resonator system.
The quantities $\delta = \omega_a - \omega$ and $\Delta_b = \omega_b - \omega_a$ are the unperturbed drive detuning and resonator detuning, respectively.
The resulting steady-state field in the main cavity is given by
\begin{equation}
    \label{eq:field_amplitude}
    \frac{A}{S_+} = \frac{\sqrt{\kappa}}{i\delta_{\text{eff}} + \Gamma_{\text{eff}}/2} = \frac{\sqrt{\kappa}\left[ i \left( \Delta_b + \delta \right) + \Gamma_b/2 \right]}{\left( i\delta + \Gamma_a/2 \right) \left[i \left( \Delta_b +\delta \right) + \Gamma_b/2 \right] + |g|^2/4}.
\end{equation}
This can be compared to the maximum steady-state field in the absence of the auxiliary resonator $A_0/S_+ = 2\sqrt{\kappa}/\Gamma_a$ as obtained by setting $|g|=0$ and $\omega = \omega_a$.
The resulting normalized steady-state field $|A|/|A_0|$ is shown in Fig.~\ref{fig:two_rings_model}a as a function of the drive and resonator detunings ($\delta,\Delta_b$).
We observe characteristic anti-crossing behavior with field maxima closely tracking the eigenvalues of an equivalent lossless system for $|g|\gtrsim \Gamma_a,\Gamma_b$.
For large resonator detunings $|\Delta|\gg|g|$, the normalized steady-state field at $\delta = 0$ approaches unity and is largely insensitive to changes in $\Delta$ indicating decoupling of the auxiliary and main resonator modes.
This demonstrates that the tuning mechanism can affect a targeted mode in the main resonator without significantly perturbing other modes, provided significant detuning is maintained.

This \textit{effective-resonator} picture shows explicitly that the auxiliary frequency $\omega_b$ (and consequently resonator detuning $\Delta_b$) can be tuned to reduce the effective detuning $\delta_{\text{eff}}$ at the expense of introducing additional effective loss $\Gamma_{\text{eff}}\geq \Gamma_a$.
These competing effects must be balanced to optimize the particular performance metric under consideration.
In many cases, maximization of the steady-state resonator field Eq.~\eqref{eq:field_amplitude} is desired which depends on minimizing $|i\delta_{\text{eff}}+\Gamma_{\text{eff}}/2|$.
We determine the optimal resonator detuning in this case to be given by
\begin{equation}
    \label{eq:optimal_detuning}
    \Delta_b^{\text{opt}} =  -\delta + \frac{|g|^2+2\Gamma_a\Gamma_b}{8\delta} + \frac{1}{2} \sqrt{\frac{|g|^4+4\Gamma_a\Gamma_b|g|^2 + 4\Gamma_a^2\Gamma_b^2}{16\delta^2} + \Gamma_b^2}.
\end{equation}
This relation can be used to find the required tuning range for the auxiliary resonator in order to implement a desired tuning range for the main resonator.
In the limit of $|g|\gg\Gamma_a,\Gamma_b$, the optimal detuning can be simplified to $\Delta_b^{\text{opt}} \approx -\delta + |g|^2/4\delta$.

The normalized cavity field $|A/A_0|$ along the $\Delta_b^\mathrm{opt}$ contour is plotted as a function of the drive detuning $\delta$ in Fig.~\ref{fig:two_rings_model}b.
The field roughly follows a Lorentzian lineshape with full-width-at-half-max $\approx\sqrt{\Gamma_a/\Gamma_b}$.
Explicit calculation of the normalized steady-state energy $|A/A_0|^2$ reveals a 3-dB tuning ($\delta$) bandwidth of
\begin{equation}
    \label{eq:tuning_bandwidth}
    B_{\delta} = \sqrt{(\sqrt{2}-1)|g|^2\frac{\Gamma_a}{\Gamma_b} + \Gamma_a^2},
\end{equation}
which is plotted against values extracted from the field directly in Fig.~\ref{fig:two_rings_model}c.
We observe that the tuning bandwidth can be made arbitrarily large provided $|\Gamma_a|\gg|\Gamma_b|$ --- even if the inter-cavity coupling is relatively weak ($|g|\approx\Gamma_a$).
This could be achieved, for example, in a system where the inter-cavity coupling is large compared to the intrinsic losses ($|g|\gg \gamma_a,\gamma_b$) but the main resonator quality factor is intentionally spoiled by significant over-coupling to the waveguide (i.e. by designing coupling $\kappa \gg \gamma_a,\gamma_b$ so that $\kappa \approx \Gamma_a \gg \Gamma_b$).
Beyond the potential to tune the resonance over several linewidths $\Gamma_a$, the proposed operation in the over-coupled regime $\kappa\approx\Gamma_a$ is also desirable in many nonlinear processes as will be discussed in Section~\ref{sec:nonlinear}.
The CMT model and its predicted effects are verified using finite-difference-time-domain simulation in Appendix~\ref{sec:numerical_validation}.

\section{Nonlinear processes in auxiliary-resonator-tuned systems} \label{sec:nonlinear}
To generalize the auxiliary resonator model for applications involving nonlinear mode interactions, a resonator configuration similar to Fig.~\ref{fig:model}b is considered.
A single main resonator supports multiple modes $a_j$ at frequencies $\omega_{a_j}$, each of which may be coupled to an associated auxiliary resonator mode $b_j$ and an input/output mode $s^\pm_{j}$.
The main resonator modes can interact through nonlinear functions $N_j(\vb a)$ corresponding to the phase-matched resonant process specifically targeted by the design.
In contrast, the auxiliary resonator design(s) will generally not support phase-matched multi-resonant nonlinear processes.
Thus, the system can be described with a series of equation pairs:
\begin{subequations}
    \begin{align}
        \label{eq:general_nonlinear_1}
        \dv{a_j}{t} &= -i\left( \omega_{a_j} - i\frac{\Gamma_{a_j}}{2} \right) a_j - i \frac{g_j^*}{2} b_j + \sqrt{\kappa_j} s_{j}^+ + N_j(\vb a), \\
        \label{eq:general_nonlinear_2}
        \dv{b_j}{t} &= -i\left( \omega_{b_j} - i\frac{\Gamma_{b_j}}{2} \right) b_j - i \frac{g_j}{2} a_j.
    \end{align}
\end{subequations}

For a given nonlinear process the various modes will be driven, either externally or by the nonlinear interaction, at a single frequency $\omega_{d,j}$.
The steady-state response is taken to be $a_j(t) = A_j\exp(-i\omega_{d,j} t)$ and $b_j(t) = B_j\exp(-i\omega_{d,j} t)$.
This enables an effective-resonator description for each pair of modes $j$, 
\begin{equation}
    0 = -\left( i\delta_{j,\text{eff}} + \frac{\Gamma_{j,\text{eff}}}{2} \right) A_j + \sqrt{\kappa_j} S_{j}^+ + N_j(\vb A) 
\end{equation}
where $S_{j}^+$ is the waveguide field amplitude at $\omega_{d,j}$ and $(\delta_{j,\text{eff}},\Gamma_{j,\text{eff}})$ are given by Eqs.~\eqref{eq:effective_eval_1} and \eqref{eq:effective_eval_2} as in the single-frequency case.

\subsection{Difference-frequency generation}

This section illustrates the effect of auxiliary resonator tuning on a nonlinear process using the specific case of resonantly-enhanced $\chi^{(2)}$ difference frequency generation (DFG), which is of particular interest for quantum information applications such as long-distance entanglement distribution over silica optical fiber.
In the multi-resonant DFG process ($\omega_i \leftrightarrow \omega_p + \omega_o$) shown in Fig.~\ref{fig:model}b, the main resonator supports input $a_1$, pump $a_2$, and output $a_3$ modes at frequencies $\omega_1 \approx \omega_i$, $\omega_2 \approx \omega_p$, and $\omega_3 \approx \omega_o$.
The input and pump resonant modes are driven by incoming waveguide modes $s^+_{1}$ and $s^+_{2}$, while all three resonator modes are coupled to outgoing waveguide modes $s^-_j$.
Two separate auxiliary rings $b_2$ and $c_3$ have resonances $\omega_b \approx \omega_2$ and $\omega_c \approx \omega_3$ and are coupled to $a_2$ and $a_3$ with rates $g_b$ and $g_c$ respectively.
The system dynamics are described by the CMT equations
\begin{subequations}
    \begin{align}
        \dv{a_1}{t} &= - i \left(\omega_1 - i \frac{\Gamma_1}{2} \right)a_1 - i \omega_1 \beta a_2 a_3 + \sqrt{\kappa_1} s^+_1, \\
        \dv{a_2}{t} &= - i \left(\omega_2 - i \frac{\Gamma_2}{2} \right)a_2 - i \omega_2 \beta^* a_1 a_3^* - i \frac{g_b^*}{2} b_2 + \sqrt{\kappa_2} s^+_2, \\
        \dv{a_3}{t} &= - i \left(\omega_3 - i \frac{\Gamma_3}{2} \right)a_3 - i \omega_3 \beta^* a_1 a_2^* - i \frac{g_c^*}{2} c_3, \\ 
        \dv{b_2}{t} &= - i \left(\omega_b - i \frac{\Gamma_b}{2} \right)b_2 - i \frac{g_b}{2} a_2, \\
        \dv{c_3}{t} &= - i \left(\omega_c - i \frac{\Gamma_c}{2} \right)c_3 - i \frac{g_c}{2} a_3, \\
        s^-_j &= - s^+_j + \sqrt{\kappa_j} a_j,
    \end{align}
\end{subequations}
where $\beta$ is a nonlinear mode overlap coefficient which encompasses effects from phase matching and spatial field overlap~\cite{burgess2009difference, logan2023sfg}.
For this analysis, $\beta$ is treated as constant, which is valid for detunings much less than a free spectral range.
The effects of detuning and auxiliary resonators on phase matching in a ring resonator geometry are detailed in Appendix~\ref{sec:phase_matching}.
Auxiliary resonators are not used to tune the highest-frequency mode $\omega_1$ because reaching similar coupling rates with the shorter wavelength would require smaller distances between the main and auxiliary resonators compared to the auxiliary resonators for $\omega_2$ or $\omega_3$.
The smaller intercavity distance could induce excessive scattering losses in the longer-wavelength modes.
Instead, all three modes would be simultaneously shifted (e.g. by temperature tuning of the main resonator) to achieve resonance $\omega_1 = \omega_i$, and then the auxiliary resonators $b_2$ and $c_3$ are tuned to shift $\omega_2$ and $\omega_3$ onto the desired output $\omega_o$ and pump $\omega_p = \omega_i - \omega_p$ frequencies respectively.

The steady-state fields in the effective-resonator picture satisfy
\begin{subequations}
\begin{align}
    \label{eq:dfg_effective_1}
    0 &= -\left(i\delta_1 + \frac{\Gamma_1}{2} \right)A_1 - i \omega_1 \beta A_2 A_3 + \sqrt{\kappa_1} S^+_{1}, \\
    \label{eq:dfg_effective_2}
    0 &= -\left(i\delta_{2,\text{eff}} + \frac{\Gamma_{2,\text{eff}}}{2} \right)A_2 - i \omega_2 \beta^* A_1 A_3^* + \sqrt{\kappa_2} S^+_{2}, \\
    \label{eq:dfg_effective_3}
    0 &= -\left(i\delta_{3,\text{eff}} + \frac{\Gamma_{3,\text{eff}}}{2} \right)A_3 - i \omega_3 \beta^* A_1 A_2^*,
\end{align}
\end{subequations}
with $S^+_{1}$ and $S^+_{2}$ as the input and pump driving amplitudes.
Throughout this section, the drive detunings are labeled $\delta_1 = \omega_1 - \omega_i$, $\delta_2 = \omega_2 - \omega_p$, and $\delta_3 = \omega_3 - \omega_o$, and the resonator detunings are $\Delta_b = \omega_b - \omega_2$ and $\Delta_c = \omega_c - \omega_3$.

In the small-signal limit of the DFG process, the converted light in the output-frequency resonator mode is negligible compared to either of the externally-driven modes.
In this case, the small-signal conversion efficiency $\eta_{ss}$ can be derived by solving the steady-state equations while neglecting the nonlinear ($\beta$) terms in Eqs.~\eqref{eq:dfg_effective_1} and \eqref{eq:dfg_effective_2}:
\begin{equation}
    \eta_{ss} \equiv \abs{\frac{S^-_{3}}{S^+_{1}S^+_{2}}}^2 = \frac{\omega_3^2|\beta|^2\kappa_1\kappa_2\kappa_3}{|i\delta_1+\Gamma_1/2|^2\,|i\delta_{2,\text{eff}}+\Gamma_{2,\text{eff}}/2|^2\,|i\delta_{3,\text{eff}}+\Gamma_{3,\text{eff}}/2|^2}.
\end{equation}
The small signal efficiency is maximized when the steady-state field of each participant mode is independently maximized, so the same choice of resonator detuning in Eq.~\eqref{eq:optimal_detuning} remains optimal.
Consequently, the modification of the small-signal efficiency and corresponding bandwidth match that of the single-resonator model Eq.~\eqref{eq:tuning_bandwidth}.
Similar to a triple-resonant process in a single resonator, $\eta_{ss}$ is maximized when all three main resonator modes are effectively critically coupled to the waveguide: $\kappa_j = \Gamma_{j,\mathrm{eff}}/2$.
However, the optimal waveguide coupling for auxiliary-tuned modes can vary depending on $\Gamma_{\mathrm{eff}}$, which increases with the magnitude of the corrected detuning.

The quantum-limited maximum conversion efficiency can be investigated using the undepleted-pump approximation in which $|S^+_{1}|\ll|S^+_{2}|$.
In this regime, only the nonlinear term in the pump mode Eq.~\eqref{eq:dfg_effective_2} can be neglected, resulting in an effective coupling between the input and output modes
\begin{equation}
    g_{\text{NL}} = \omega_1\beta A_2 = \omega_1\beta\frac{\sqrt{\kappa_2} S^+_{2}}{i\delta_{2,\text{eff}} + \Gamma_{2,\text{eff}}/2}
\end{equation}
Under the assumption that $\delta_1 = 0$ by global tuning of the main resonator, the conversion efficiency is given by
\begin{equation}
    \eta = \frac{r^2|g_{\text{NL}}|^2 \kappa_1 \kappa_3}{\big( (\Gamma_1\Gamma_{3,\text{eff}}/4)+r|g_{\text{NL}}|^2 \big)^2 + \delta_{3,\text{eff}}^2\Gamma_1^2/4}
\end{equation}
where $r = \omega_3/\omega_1$.
Optimal conversion efficiency occurs for pump powers
\begin{equation}
    |S^+_{2\text{, opt}}|^2 = \frac{\Gamma_1}{2|\beta|^2\omega_1\omega_3\kappa_2} |\delta_{2,\text{eff}} + i \Gamma_{2,\text{eff}}/2|^2 |\delta_{3,\text{eff}} + i \Gamma_{3,\text{eff}}/2|
\end{equation}
which is similarly minimized when the resonator detuning is given by Eq.~\eqref{eq:optimal_detuning}.
The corresponding optimal conversion efficiency is then
\begin{equation}
    \label{eq:dfg_optimal_efficiency}
    \eta_{\text{opt}} = \frac{2\omega_3\kappa_1\kappa_3}{\omega_1\Gamma_1 \Big( 2|\delta_{3,\text{eff}} + i\Gamma_{3,\text{eff}}/2| + \Gamma_{3,\text{eff}} \Big)}.
\end{equation}
We find that $\eta_{\mathrm{opt}}$ is approximately maximized for resonator detunings which maximize the cavity fields Eq.~\eqref{eq:optimal_detuning}.
The resulting $\eta_{\mathrm{opt}}$ can be compared to the ideal system quantum limit \cite{burgess2009difference}
\begin{equation}
    \eta_{\text{QL}} = \frac{\omega_3 \kappa_1 \kappa_3}{\omega_1 \Gamma_1 \Gamma_3}.
\end{equation}
which corresponds to $\delta_{3,\text{eff}} = 0$ and $\Gamma_{3,\text{eff}}=\Gamma_3$.

\section{Practical tuning range and bandwidth} \label{sec:projected_performance}

\begin{figure}[t]
    \centering
    \includegraphics[width=\textwidth]{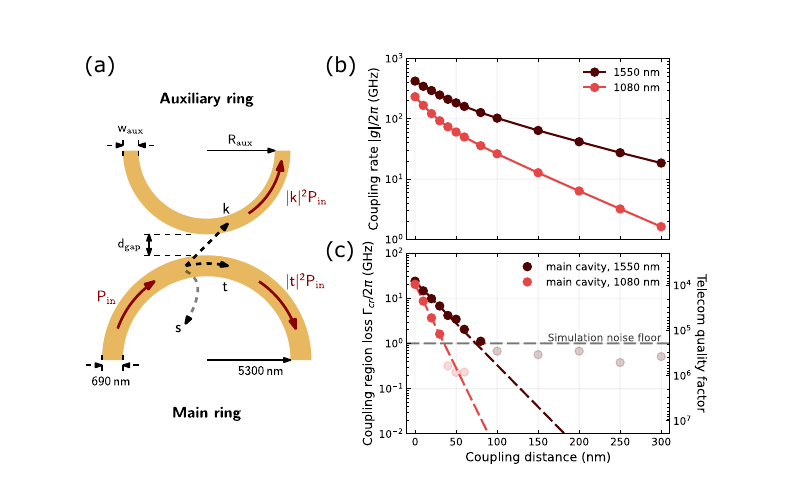}
    \caption{(a) Simulation diagram of the coupling region between two rings used to simulate single-pass transmission and inter-ring coupling. 
    (b) Dependence of inter-cavity coupling rate on inter-cavity distance, derived from simulated single-pass coupling, for an auxiliary resonator design with $R_{\text{aux}}=3$\,\textmu m and $w_{\text{aux}}=400$\,nm.
    The coupling $k$, transmission $t$, and scattering $s$ fractions refer to the field amplitude.
    (c) Excess scattering loss rates in the main cavity telecom and pump band modes induced by the coupled auxiliary resonators.
    For the telecom mode, equivalent scattering quality factor is shown on the right-side axis.
    Loss rates below $\mathrm{\sim 1\,GHz}$ at 1550\,nm cannot be accurately estimated by varFDTD due to single-pass power loss dropping below the simulation noise floor.
    Exponential fits to the remaining points indicate a reduction in loss of about 2 and 3.5 orders of magnitude every 100\,nm for 1550\,nm and 1080\,nm respectively.
    } 
    \label{fig:single_pass}
\end{figure}

The above analysis demonstrates the potential of the auxiliary resonator tuning mechanism in enabling efficient and targeted multi-resonant nonlinear processes.
The bandwidth over which such processes can be achieved, however, is largely determined by the inter-cavity coupling rate $|g|$ (Eq.~\eqref{eq:tuning_bandwidth}).
Consequently, the practicality of the tuning scheme depends on the range of experimentally obtainable coupling rates in realistic device designs.
In this section, we simulate both the coupling rates and added scattering losses resulting from an auxiliary ring resonator coupled to a triply-resonant ring resonator design \cite{logan2023sfg}.
Based on these simulations, we then examine a resonant DFG process at targeted frequencies in the combined system in order to characterize the tuning bandwidth and performance.

\subsection{Coupling rates and losses}
To determine $|g|$ for a ring or disk resonator, the propagation of a broadband pulse through the coupling region, depicted in Fig.~\ref{fig:single_pass}a, is simulated with a variational finite-difference time-domain method (varFDTD, Lumerical) to determine the power scattering matrix.
varFDTD performs an approximately equivalent two-dimensional simulation using effective indices derived from a three-dimensional structure.
The normalized single-pass power coupling spectrum $|k|^2$ can then be related to the inter-cavity coupling rate $|g|$ as
\begin{equation}
    \label{eq:single_pass_g}
    |g| = \frac{2|k|}{\sqrt{\tau_a\tau_b}}
\end{equation}
where $\tau_x$ is the round-trip time of the cavity~\cite{little_jlightwavetech1997_singlepass}.
For this particular configuration $\tau_x = v_{g,x} / 2\pi R_x$ with $v_{g,x}$ and $R_x$ as the group velocity and radius of the resonators, respectively.
However, the presence of the auxiliary ring simultaneously introduces scattering losses at the coupling region due to the perturbation of the evanescent field.  
This excess scattering loss at the coupling region $|s|^2$ is similarly calculated from the single-pass varFDTD simulations by comparing the combined transmitted $|t|^2$ and coupled $|k|^2$ powers to the transmission in the absence of the auxiliary ring.
The corresponding coupling region loss rate is given by $\Gamma_{cr,x} = |s|^2/\tau_x$.
It is expected that both the coupling rate and scattering loss depend on the evanescent field overlap and thus decay exponentially with increasing separations $d_{\text{gap}}$.

The main resonator geometry is based on the design from Ref.~\cite{logan2023sfg}, which utilizes a hybrid-integrated gallium phosphide(GaP)-on-oxide(SiO$_2$) ridge-waveguide ring of inner radius of 5.3\,\textmu m, width of 690\,nm, and height of 430\,nm to target phase-matched resonances at 637\,nm, 1080\,nm, and 1550\,nm for DFG.
A range of auxiliary resonator radii $R_{\text{aux}}$, widths $w_{\text{aux}}$, and separations $d_{\text{gap}}$ were simulated ranging from 3--8\,\textmu m, 200--500\,nm, and 0--300\,nm respectively.
Frequency-domain simulations (Lumerical MODE) are utilized to determine the group indices and bending losses for propagating modes near 1080\,nm and 1550\,nm.
This enables the determination of the inter-cavity coupling rates via Eq.~\eqref{eq:single_pass_g}.

The coupling rate $|g|$ most strongly depends on the inter-cavity separation distance $d_{\text{gap}}$, as shown in Fig.~\ref{fig:single_pass}b. 
The dependence of $|g|$ on the auxiliary ring geometry is comparatively weak (Appendix~\ref{sec:single_pass_coupling}).
In general, the coupling rate increases for smaller ring diameters with shorter round-trip propagation times and smaller ring widths with stronger evanescent fields, so long as the auxiliary resonator mode is not approaching cutoff.
As expected, we observe a roughly exponential decay of the coupling rate $|g|$ with increasing inter-cavity separation.
For small separations ($d_{\text{gap}} < 50$\,nm), the auxiliary resonator begins to strongly perturb the main resonator modes causing increased coupling rates but also increased scattering losses $\Gamma_{cr}$ (Fig.~\ref{fig:single_pass}c).
Consequently, the optimal coupling distance corresponds to the smallest distance for which the scattering losses do not significantly contribute to the total quality factor.
For a design targeting loaded quality factors of $Q=10^5$, a coupling distance of $d_{\text{gap}} = 125$\,nm would provide a sufficiently high telecom coupling rate ($|g|\approx 90$\,GHz) while introducing relatively minor losses ($Q_{cr} \approx 10^6$).

\subsection{Difference-frequency generation}

\begin{table}[bt]
    \centering
    \begin{tabular}{c | c c c c c c c }
        mode & $\lambda_0$ (nm) & $Q_j$ & $\Gamma_{j}/2\pi$ & $Q_{b_j}$ & $\Gamma_{b_j}/2\pi$  & $|g_j|/2\pi$  & $B_{\delta,j}/\Gamma_j$ \\
        \hline
         1 & 637 
            & $9.1\!\times\!10^4$
            & 5.18
            & --
            & --
            & --
            & --
         \\
         2 & 1080 
            & $9.1\!\times\!10^4$
            & 3.05
            & $1\!\times\!10^6$
            & 0.28
            & 15
            & 11
         \\
         3 & 1550 
            & $9.1\!\times\!10^4$
            & 2.13
            & $1\!\times\!10^6$
            & 0.19
            & 64
            & 64
    \end{tabular}
    \caption{Parameters for DFG performance estimation.
    All rates are in GHz.}
    \label{tab:dfg_params}
\end{table}

\begin{figure}[t]
    \centering
    \includegraphics[width=\textwidth]{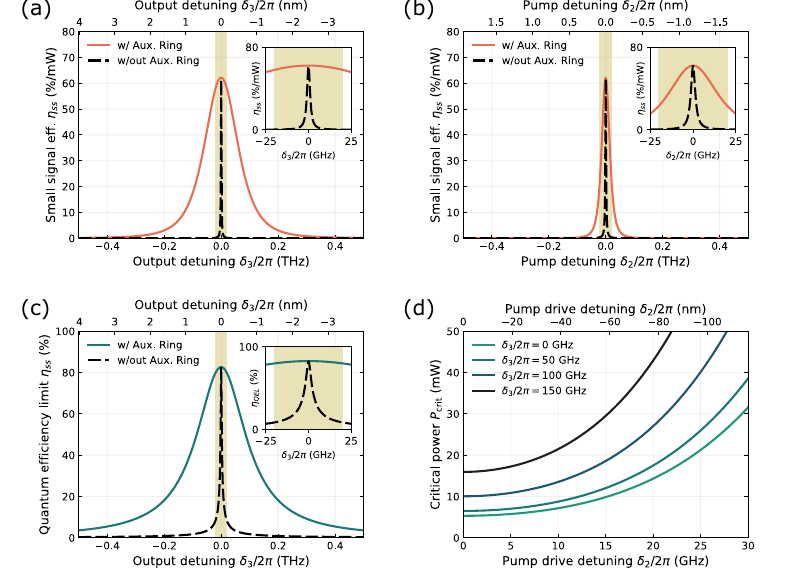}
    \caption{Small-signal photon conversion efficiency with and without an auxiliary resonator as a function of (a) output and (b) pump mode detuning.
    All detunings are measured relative to the unperturbed (sans auxiliary resonator) main resonator frequencies ($\omega_2$, $\omega_3$).
    Insets show the indicated regions with re-scaled detuning axis.
    (c) Quantum-efficiency-limited (QEL) photon conversion efficiency as a function of the output drive detuning $\delta_3$.
    Pump mode detunings do not affect the value of $\eta_{\text{QEL}}$ but are shown to modify the critical power in (d).
    Critical values are shown to remain in the few mW range over the tuning bandwidths of both pump $\delta_2$ and output $\delta_3$ modes.
    }
    \label{fig:dfg_performance}
\end{figure}

We now model the performance of a triply resonant DFG process (based on \cite{logan2023sfg}) with the addition of auxiliary ring resonators. 
We consider an auxiliary ring design with nominal dimensions $R_{\text{aux}}=3$\,\textmu m, $w_{\text{aux}}=400$\,nm, and $d_{\text{gap}}=150$\,nm.
The resulting pump (1080\,nm) and output (1550\,nm) inter-cavity coupling rates are $|g_b|/2\pi\approx 15$\,GHz and $|g_c|/2\pi\approx64$\,GHz respectively.
We assume all three modes have fabrication-limited intrinsic quality factors of $Q_i = 10^6$ (quality factors in GaP photonics at visible/telecom wavelengths have been observed to be well above $10^5$ \cite{logan2023sfg, yama_advmat2023_bgap, wilson2020gap_comb}).
Coupling-region loss at the chosen $d_{\text{gap}} = 150$\,nm corresponds to scattering quality factors far exceeding $10^6$ (Fig.~\ref{fig:single_pass}c) and may consequently be neglected.

Coupling to the waveguide introduces an additional loss channel within the main resonator but not the auxiliary resonators.
We assume that the main resonator is significantly over-coupled, corresponding to a coupling quality factor of $Q_c=10^5$ for all three modes.
The resulting total quality factors for the three main resonator modes are then $Q_j = 9.1\times10^4$ ($j=1,2,3$).
The auxiliary rings do not couple to the waveguide and so $Q_b = Q_c = 10^6$.
We then compute the 3-dB tuning bandwidth Eq.~\eqref{eq:tuning_bandwidth} for the pump and output modes to be
\begin{align*}
    B_{\delta,2}/2\pi &\approx 32\,\text{GHz} \approx 11 (\Gamma_2 / 2\pi), \\
    B_{\delta,3}/2\pi &\approx 136\,\text{GHz} \approx 64 (\Gamma_3 / 2\pi).
\end{align*}
The assumed resonator parameters and the corresponding bandwidths are summarized in Table~\ref{tab:dfg_params}.
As described in Eq.~\eqref{eq:optimal_detuning}, the accessible portion of this tuning curve is limited by the tuning range of the auxiliary resonator.
Also, if a resonance is tuned by more than approximately 20\% of a free spectral range, the conversion efficiency may be reduced due to phase matching effects (Appendix~\ref{sec:phase_matching}), though this is not likely to become significant in high-finesse systems.

Finally, we compute the projected DFG performance metrics as shown in Fig.~\ref{fig:dfg_performance} assuming the input mode has been tuned globally onto resonance (i.e. $\delta_1=0$).
We observe that the pump and output modes could be tuned over a range of 4 and 23 main-resonator linewidths, respectively, while maintaining over 90\% of the maximum small signal conversion efficiency.
The small signal efficiency can be increased at the expense of the pump mode tuning bandwidth by reducing the coupling rate $\kappa_2$ to the near-critically coupled regime ($\kappa_2 \approx \gamma_2$).
Although the tuning bandwidth of the pump mode is diminished compared to the output mode, the critical power is relatively insensitive to modulation, remaining near 10\,mW over the full range (Fig.~\ref{fig:dfg_performance}d).
Operation at the quantum efficiency limit can be achieved over the much larger telecom tuning bandwidth (Fig.~\ref{fig:dfg_performance}c) even in spite of the comparably worse pump bandwidth.
These results indicate that the auxiliary resonator tuning method could enable high-efficiency, frequency-targeted DFG over large bandwidths with minimal impact in performance.

\section{Conclusion and outlook}

Coupled auxiliary resonators offer a straightforward method to actively and independently tune specific resonances in a multi-resonant device without altering the structure of the primary resonator.
The tuning scheme can be implemented and used in conjunction with any resonator-specific tuning mechanism, allowing for compatibility with most material platforms while also limiting the required fabrication complexity. 
This technique enables the tuning of individual resonances over tens of linewidths and can be accurately modeled by a small number of fast and computationally inexpensive simulations.

The flexibility and control uniquely afforded by the auxiliary-resonator tuning scheme is particularly relevant to quantum information applications which impose strict restrictions on process wavelengths.
For example, difference-frequency conversion of photons emitted from solid-state qubits---such as the nitrogen-vacancy \cite{dreau_prappl2018_nvtelecom} and silicon-vacancy \cite{bersin_arxiv2023_sivtelecom} centers in diamond---to a specific telecom wavelength could be used to not only minimize losses on a fiber-based quantum network \cite{weber2019twophoton}, but also correct for spectral inhomogeneity of the emitter nodes.
In this way, the high conversion efficiency afforded by the multi-resonant nonlinear process, combined with the flexibility of the auxiliary resonator tuning mechanism, may serve an indispensable role in the development of large-scale quantum networks.

\section{Acknowledgement}
This material is based upon work supported by Department of Energy, Office of Science, National Quantum Information Science Research Centers, Co-design Center for Quantum Advantage (C2QA) under contract number DE-SC0012704 and National Science Foundation Grant No. ECCS-1807566. N.S.Y.\ was supported by the National Science Foundation Graduate Research Fellowship Program under Grant No.~DGE-2140004.

\section{Disclosures}
The authors declare no conflicts of interest.

\section{Data Availability}
The simulation data and calculations in this work can be obtained by contacting the authors.

\appendix

\section{Numerical validation} \label{sec:numerical_validation}
\begin{figure}[h]
    \centering
    \includegraphics[width=\textwidth]{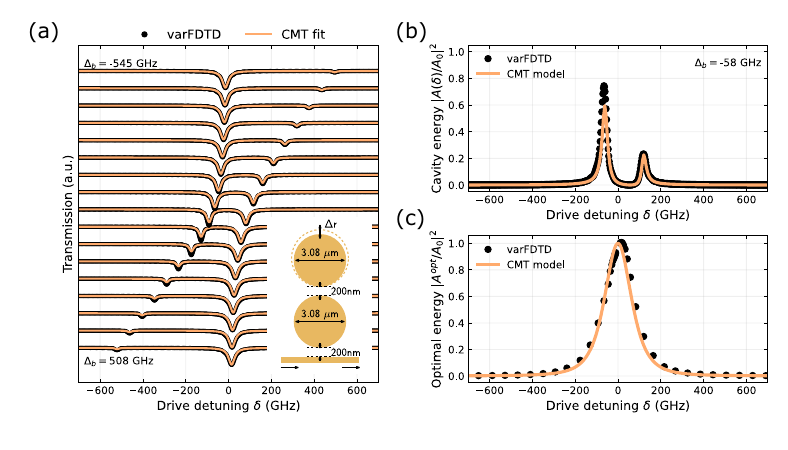}
    \caption{
    (a) Transmission spectra obtained from varFDTD and CMT model fits for an auxiliary-resonator system (inset) at different resonator detunings.
    The coupling distances are held constant while the auxiliary resonator radius is modulated to detune the cavities.
    (b) Simulated and predicted energy in the main resonator for $\Delta_b = -58$\,GHz.
    (c) Simulated and predicted theoretical maximum energies in the cavity as a function of drive detuning.
    Throughout the figure, the CMT model curves are determined by coupling and loss rates ($g$, $\gamma$, and $\kappa$) from single pass simulations (Section 4) and simulations of the structure without the auxiliary resonator.
    Only the relative mode frequencies were adjusted to fit the FDTD data in order to compensate for slight perturbations from single-resonator simulations.
    }
    \label{fig:numerical_validation}
\end{figure}

To verify the predictions from the CMT model, we simulate a GaP-on-oxide coupled-auxiliary-resonator system using varFDTD, which performs an approximately equivalent two-dimensional simulation using effective indices derived from the three-dimensional structure.
For the highly-planar devices under consideration, varFDTD enables reasonably accurate simulation of device performance with much higher throughput.
The system, shown schematically in the inset of Fig.\ref{fig:numerical_validation}a, consists of two GaP disk resonators on a silicon-oxide substrate.
One resonator is evanescently coupled to a waveguide, which is used to excite the resonators.
By varying the auxiliary resonator radius while maintaining the edge-to-edge separation of the resonators, the auxiliary resonator detuning $\Delta_b$ can be modulated without significantly affecting the coupling rate $g$ directly.

Accurate simulation of steady-state device transmission and cavity fields requires the transient response to completely decay over the course of the simulation.
To achieve this while minimizing computational resources, we use a comparatively small geometry with nominal radii of $1.54$\,\textmu m and a height of 500\,nm.
Furthermore, intrinsic material losses within the resonators ($\Im(n_{\mathrm{GaP}}) = 10^{-4}$) are imposed, which has an effect similar to scattering losses from surface roughness in fabricated devices.
Simulations of the main resonator transmission in the absence of the auxiliary resonator reveal an intrinsic loss rate $\gamma/2\pi = 20.4$\,GHz and waveguide coupling loss rate $\kappa/2\pi = 14.6$\,GHz.
By varying the radius of the auxiliary resonator by $\pm 5.5$\,nm in $0.5$\,nm steps, the frequency of the auxiliary resonance is tuned by approximately $\pm 750$\,GHz.
Single-pass coupling simulations (discussed in main text Section 4) predict coupling rates of $g/2\pi \approx 170$\,GHz.

Simulated transmission spectra are shown in Fig.~\ref{fig:numerical_validation}a.
The CMT model is fit using known values of the loss and coupling rates ($\gamma$, $\kappa$, and $g$) obtained from single-pass simulations (main text, Section 4) with the only free parameters being the absolute and relative frequencies of the resonators.
In Fig.~\ref{fig:numerical_validation}b, the normalized steady-state main cavity energy obtained from one of these simulations is compared to the corresponding CMT prediction.
Fig.~\ref{fig:numerical_validation}c shows the highest attainable cavity energy as a function of drive detuning $\delta$ compared to the unperturbed resonance, under optimal auxiliary resonator detuning $\Delta_b$.
The simulation results closely follow the Lorentzian shape of the CMT model curve, which is identical to that in Fig.~2b except using the known parameters.
In each of these plots, reasonably good agreement is observed between the varFDTD simulation results and CMT model predictions, indicating that the model can provide useful predictions of system behavior using only a few parameters which can be extracted from a computationally inexpensive set of simulations.

\section{Single-pass coupling simulations}
\label{sec:single_pass_coupling}
\begin{figure}[h]
    \centering
    \includegraphics[width=\textwidth]{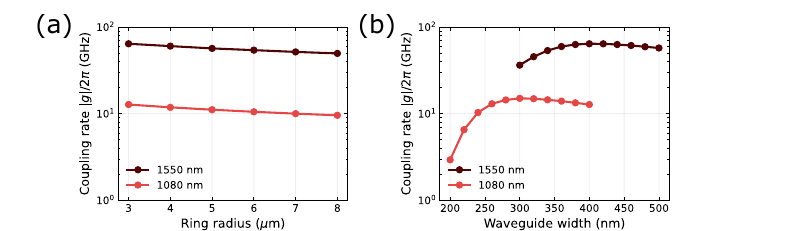}
    \caption{Dependence of the coupling rate $g$ as determined by single-pass coupling simulations as a function of auxiliary ring resonator (a) radius and (b) width.
    For both simulations the coupling distance $d_{\mathrm{gap}} = 150$\,nm and (a) takes the width to be $w_{\mathrm{aux}}=400$\,nm while (b) assumes a radius of $r_{\mathrm{aux}}=3.0$\,\textmu m.
    }
    \label{fig:single_pass_apdx}
\end{figure}

Simulations of the single pass coupling and loss were performed for the full, multi-dimensional parameter sweep.
Representative behavior is shown in Fig.~\ref{fig:single_pass_apdx} for variation with respect to the auxiliary ring resonator radius and width.
In general these parameters do not strongly affect the coupling rate provided the width and radius is sufficiently large as to allow for low-loss propagation of the auxiliary resonator mode.
As expected, the strongest dependence is on the separation between the two rings.

\section{Phase matching in ring resonators} \label{sec:phase_matching}
\begin{figure}[h]
    \centering
    \includegraphics[width=\textwidth]{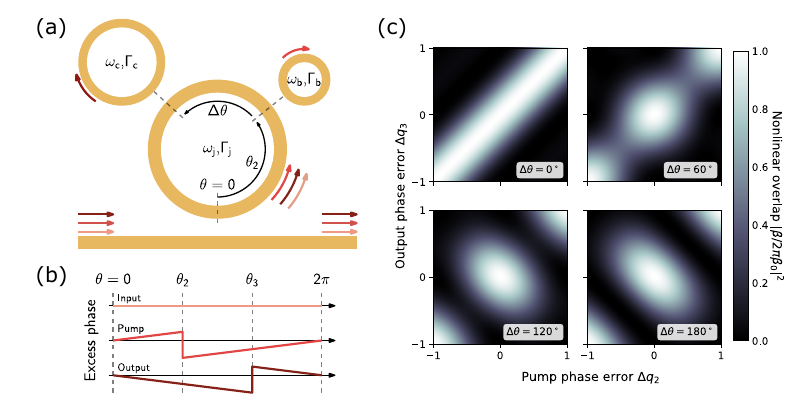}
    \caption{(a) Schematic of a DFG process $\omega_1 \leftrightarrow \omega_2 + \omega_3$ in a waveguide-coupled ring resonator, with two auxiliary resonators for tuning. The auxiliary ring resonators coupled to modes 2 and 3 are respectively located at angular positions $\theta_2$ and $\theta_3 = \theta_2 + \Delta\theta$ with respect to the main ring.
    (b) Illustration of the excess phase over the circumference of the main ring resonator.
    For the input field phase, the resonant driving ensures that no excess phase is incurred whereas the pump and output modes are detuned and thus incur additional phase which is discontinuously corrected at $\theta_2$ and $\theta_3$ by the respective auxiliary resonators.
    This results in no net phase accumulation over a round trip.
    (c) Nonlinear overlap $|\beta|^2$ as a function of angular propagation constant mismatch $\Delta q_2$ and $\Delta q_3$ for different separations of the auxiliary resonators $\Delta\theta$.
    }
    \label{fig:phase_matching}
\end{figure}

For nonlinear processes in a ring or whispering gallery resonator, the angular propagation constants $q_j$ describe the spatial phase evolution of each propagating mode $j$ over the resonator circumference.
At the resonant frequencies $\omega_j$, the propagation constants $q_j$ are integers equal to the azimuthal mode numbers $m_j$.
Conservation of momentum for a perfectly phase-matched nonlinear process requires the sum of propagation constants for the input and output modes to be equal.
This section uses the specific case of triple-resonant difference frequency generation $\omega_i \leftrightarrow \omega_p + \omega_o$ (Fig.~\ref{fig:phase_matching}a) in a ring resonator with resonant frequencies $\omega_1 \approx \omega_i$, $\omega_2 \approx \omega_p$, and $\omega_3 \approx \omega_o$ to illustrate phase-matching effects related to auxiliary resonator tuning.
For DFG, conversion efficiency is maximized when the phase mismatch $M \equiv q_2+q_3-q_1$ is equal to zero.
In the absence of an auxiliary resonator, the nonlinear overlap integral $\beta= \beta_0 \int_{0}^{2\pi} \dd{\theta} \exp(iM\theta)$ appearing in the CMT equations exhibits a sinc-like dependence on $M$:
\begin{equation}
    |\beta| = 2|\beta_0| \frac{\sin(\pi M)}{M}.
\end{equation}
Consequently, $|\beta|$ decreases for $M\neq0$ and vanishes completely for nonzero integer values of $M$.
Quasi-phase matched processes exhibit a similar dependence, but offset to a nonzero optimal $M$.

A multi-resonant nonlinear cavity with modes detuned from the designed process frequencies $\omega_j$ will have angular propagation constants similarly displaced from the intended integer azimuthal mode numbers: $q_j = m_j + \Delta q_j$, resulting in a round-trip phase shift of $2 \pi \Delta q_j$.
A coupled auxiliary resonator can tune a main resonator mode onto the design frequency by introducing a compensating phase shift of $-2 \pi \Delta q_j$ over the length of the intercavity coupling region, allowing constructive self-interference.
The angular propagation constant of the mode is not affected outside of the coupling region.
For a ring or disk resonator with sufficiently large diameter, the effect of the coupling region can be approximated as a phase discontinutiy at the angular location of the auxiliary resonator $\theta_j$ so that the resonant field has spatial phase
\begin{equation}
    \phi_j(\theta) = 
    \begin{cases}
        (m_j + \Delta q_j) \theta, & 0 \leq \theta < \theta_j \\
        (m_j + \Delta q_j) \theta - 2\pi\Delta q_j , & \theta_j \leq \theta < 2\pi \\
    \end{cases}.
\end{equation}
The effect of this phase discontinuity is illustrated in Fig.~\ref{fig:phase_matching}b.

An example tuning scheme is shown in Fig.~\ref{fig:phase_matching}a, consisting of a triple-resonant DFG ring resonator coupled to two auxiliary resonators.
The highest frequency mode in the main ring is tuned to the input frequency ($\omega_1 = \omega_i$) by a global tuning mechanism, while any residual detuning for the other two modes is corrected by the auxiliary resonators at angular positions $\theta_2$ and $\theta_3$.
Provided the main resonator modes satisfy phase matching ($m_1 = m_2 + m_3$), the phase mismatch of the unperturbed process is $M = \Delta q_2 - \Delta q_3$.
This yields a nonlinear overlap of
\begin{equation}
    \beta = \beta_0 \left( \int_0^{\theta_2} \dd{\theta} e^{iM \theta} 
    +
    e^{-2\pi i \Delta q_2} \int_{\theta_2}^{\theta_3} \dd{\theta} e^{iM \theta}
    +
    e^{-2\pi i (\Delta q_2 - \Delta q_3)} \int_{\theta_3}^{2\pi} \dd{\theta} e^{iM \theta}
    \right)
\end{equation}
which can be evaluated as
\begin{equation}
    |\beta| = \frac{2|\beta_0|}{|\Delta q_2 - \Delta q_3|} \left|
        \sin(\pi \Delta q_2)
        -
        e^{i(\Delta\theta -\pi)(\Delta q_2 - \Delta q_3)} \sin(\pi \Delta q_3)
    \right| 
\end{equation}
where $\Delta\theta = \theta_3 - \theta_2$.
The nonlinear overlap is plotted in Fig.~\ref{fig:phase_matching}c for $\Delta\theta = 0^{\circ}$, $60^{\circ}$, $120^{\circ}$, and $180^{\circ}$.
For most practical applications, the frequency shifts from auxiliary resonator tuning will be much less than the free spectral range of each mode ($\Delta q_j \ll 1$), resulting in negligible phase matching effects on $|\beta|$ regardless of resonator positioning.
If only one auxiliary resonator is used for tuning or if the angular separation between two auxiliary resonators is near zero, the phase matching of the tuned process remains unchanged from the unperturbed process.
Increasing the angular separation between two auxiliary resonators can allow the phase discontinuities from tuning to improve the effective phase matching of the frequency conversion process, similar to a quasi-phase matching scheme.
However, depending on the relative magnitudes and directions of the tuning applied to each mode, the effective phase matching may instead be degraded compared to the unperturbed process.

\bibliography{bib1}

\end{document}